# Programmable scanning diffuse speckle contrast imaging of cerebral blood flow

**Faezeh Akbari, [a] Xuhui Liu, [a] Fatemeh Hamedi, [a] Mehrana Mohtasebi, [a] Lei Chen, [b] Guoqiang Yu [a],***

[a]University of Kentucky, Department of Biomedical Engineering, Lexington, KY, USA
[b]University of Kentucky, Spinal Cord and Brain Injury Research Center, Department of Physiology, Lexington, KY, USA

**Abstract**

**Significance:** Cerebral blood flow (CBF) imaging is crucial for diagnosing cerebrovascular diseases. However, existing large neuroimaging techniques with high cost, low sampling rate, and poor mobility make them unsuitable for continuous and longitudinal CBF monitoring at the bedside.

**Aim:** This study aimed to develop a low-cost, portable, programmable scanning diffuse speckle contrast imaging (PS-DSCI) technology for fast, high-density, and depth-sensitive imaging of CBF in rodents.

**Approach:** The PS-DSCI employed a programmable digital micromirror device (DMD) for remote line-shape laser (785 nm) scanning on tissue surface and synchronized a 2D camera for capturing boundary diffuse laser speckle contrasts. New algorithms were developed to address deformations of line-shape scanning, thus minimizing CBF reconstruction artifacts. The PS-DSCI was examined in head-simulating phantoms and adult mice.

**Results:** The PS-DSCI enables resolving Intralipid particle flow contrasts at different tissue depths. *In vivo* experiments in adult mice demonstrated the capability of PS-DSCI to image global/regional CBF variations induced by 8% $CO_2$ inhalation and transient carotid artery ligations.

**Conclusions:** Compared to conventional point scanning, the line scanning in PS-DSCI significantly increases spatiotemporal resolution. The high sampling rate of PS-DSCI is crucial for capturing rapid CBF changes while high spatial resolution is important for visualizing brain vasculature.

**\*Corresponding Author: Guoqiang Yu, E-mail: guoqiang.yu@uky.edu

# 1 Introduction

Cerebral blood flow (CBF) plays a critical role in sustaining brain health and function. CBF serves as the conduit for delivering essential oxygen and nutrients to the brain while facilitating the removal of waste products.[1-3]. The vitality and functionality of the brain are intricately linked to the efficiency of blood circulation and hemodynamic processes. Continuous and longitudinal monitoring of CBF is crucial for understanding the pathophysiology and developing treatment strategies for many cerebral/neurovascular diseases including stroke [4-6], intraventricular hemorrhage [7, 8], and traumatic brain injury [9, 10].

Various modalities exist for cerebral hemodynamic imaging. Magnetic resonance imaging (MRI) [11-13] and positron emission tomography (PET) [14, 15] offer the capability of whole-head imaging of brain hemodynamics and metabolism. However, the high cost, low sampling rate, and poor mobility make them unsuitable for continuous and longitudinal cerebral monitoring at the bedside. Optical imaging modalities present a compelling alternative due to their safety, portability, mobility, affordability, and high temporal resolution. Laser speckle contrast imaging (LSCI) is a noncontact imaging tool with high spatiotemporal resolution but limited penetration depth (less than 1 mm) because of using wide-field illumination [16, 17]. Laser Doppler flowmetry [18, 19], diffuse correlation spectroscopy [20-22], and diffuse speckle contrast flowmetry [23-25] utilize limited discrete source-detector (S-D) pairs in contact with the head for point-of-care measurements of CBF. The sparse arrangement of S-D pairs inherently limits spatial resolutions. Increasing the number of sources and detectors escalates instrumentation costs and slows down sampling rates. Moreover, the contact measurement may be impractical for intraoperative monitoring of wounded/injured tissues.

To overcome limitations of existing technologies, researchers have prompted the development of noncontact optical imaging devices with scanning illumination and high-resolution 2D cameras for high-density imaging of blood flow distributions [26-28]. Speckle contrast diffuse correlation tomography (scDCT) [29] is one of these modalities that we have previously developed for high-density imaging of CBF distributions [29-37]. In scDCT, a galvo mirror remotely scans a near-infrared coherent point source to multiple positions on a selected region of interest (ROI) [38]. A high-resolution scientific complementary metal oxide semiconductor (sCMOS) camera, serving as a 2D detector array, captures fluctuations of spatial diffuse laser speckles resulting from motions of red blood cells in the measured tissue volume (i.e., tissue blood flow). Based on photon diffusion theories, the maximum tissue penetration depth is approximately one half of the S-D separation used [39-41]. Therefore, tissue blood flow distributions at different depths are reconstructed by quantifying spatial laser speckle contrasts in selected detector areas, defined at specific S-D separations. Continuous and longitudinal imaging of CBF distributions has been successfully demonstrated using the scDCT in rodents, piglets, and human subjects [29-36]. While effective, scDCT requires scanning of a point source to numerous positions for high-density sampling, which is time-consuming [42]. This long scanning time limits the scDCT for applications wherein a high temporal resolution is needed, such as capturing pulsatile blood flow and imaging brain functional connectivity (FC) [43-46].

To improve scanning efficiency and spatiotemporal resolution, we developed a programmable scanning diffuse speckle contrast imaging (PS-DSCI) technology for rapid and high-density imaging of CBF distributions at different tissue depths. This innovative PS-DSCI employed a programmable digital micromirror device (DMD) for remote line-shape laser scanning on the tissue surface and synchronized a high-resolution 2D camera for capturing boundary diffuse laser speckle contrasts. New algorithms were developed to analyze the collected PS-DSCI data for 2D mapping of blood flow distributions at different depths. Utilizing fast line scanning generated by the DMD, as opposed to previous point scanning in scDCT, significantly improves the spatiotemporal resolution. The PS-DSCI system was evaluated using head-simulating phantoms with known optical and flow properties. Finally, *in vivo* experiments in adult mice demonstrated the capability of PS-DSCI system to image global and regional CBF variations induced by 8% $CO_2$ inhalation and transient carotid artery ligations.

## 2 Methods

*2.1 PS-DSCI System*

*2.1.1 PS-DSCI Principle and Prototype*

In the PS-DSCI prototype (Fig. 1), an open-space linearly polarized coherent laser (785 nm, 120 mW, DL785-120-S, CrystaLaser) [38] illuminates a semi-collimated light via a biconvex lens (LB1092-B, Thorlabs) and a flat mirror (UM10-AG, Thorlabs) onto the micromirror array of a programmable DMD (DLP4500, Texas Instruments) to achieve rapid line scanning. The semi-collimated illumination provides a large depth of focus, which allows for the projection of line-shape laser scanning over the selected ROI at different working distances. The DMD enables the customizable control of preset illumination patterns, thus enhancing overall flexibility of PS-DSCI scanning. The DLP4500 comprises 1,039,680 mirrors (pitch size = 7.6 µm) arranged in a diamond pixel array geometry, which can be changed in the range of +12 or -12 degrees by varying the controlled voltage. The incident angle was optimized to align the brightest diffraction orders with the center of the energy envelope to maximize diffraction efficiency (Fig. 1a). A sCMOS camera (C11440-42U40, Hamamatsu), connected to an adjustable zoom lens (Zoom 7000, Navitar), was synchronized with the DMD to sequentially capture raw intensity images of the line scanning illumination on the ROI. A linear polarizer (LPNIRE200-B, Thorlabs) and a long-pass filter (84-761, Edmund Optics) were installed in front of the zoom lens to reduce specular reflection and ambient light, respectively. All components were installed on a vertically mounted optical breadboard to facilitate animal experiments (Fig. 1b and Fig. 1c). This compact setup allows for assembling the PS-DSCI system on a portable, movable cart to facilitate bedside measurements.

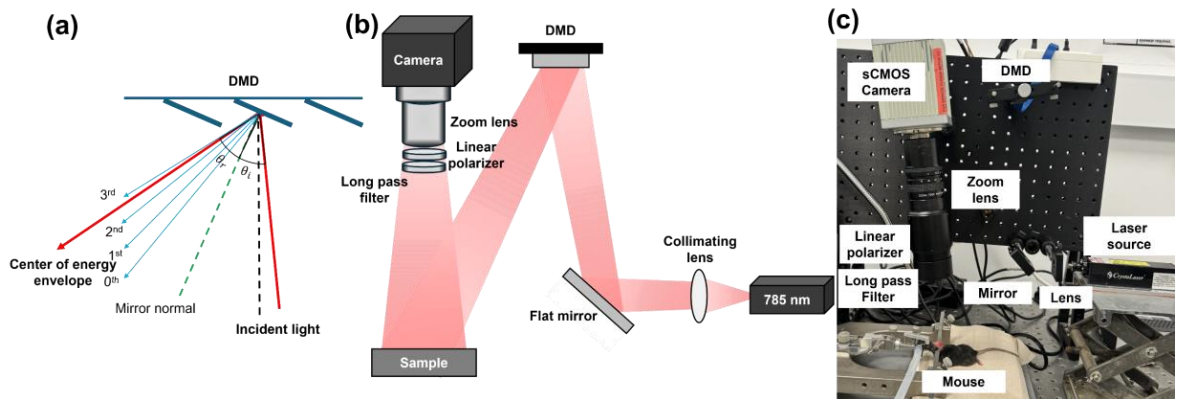

**Fig. 1** PS-DSCI system. (a) The distribution of diffraction orders and energy envelope reflected by the DMD, which are dependent on the angle of incidence ($\theta i$) and angle of reflection ($\theta r$). $\theta r$ represents the center of energy envelope distribution. (b) Schematic of the PS-DSCI prototype. (c) A photo of PS-DSCI prototype.

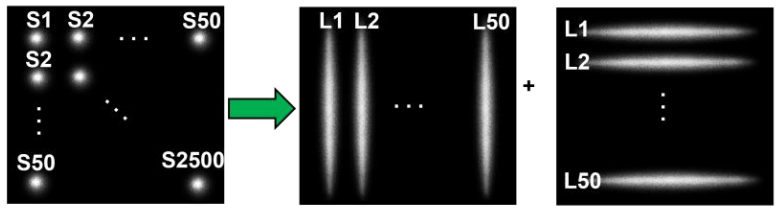

**Fig. 2** Point scanning (scDCT) versus line scanning (PS-DSCI). In contrast to 2500 scanning points (S1 to S2500) by scDCT, PS-DSCI scans the same ROI with 100 scanning lines (L1 to L50) along the vertical and horizontal directions, respectively.

*2.1.2 Transition from Point Scanning to Line Scanning*

Compared to point source scanning in scDCT, line scanning in PS-DSCI is more efficient in terms of sampling rate and number of images acquired, processed, and stored (Fig. 2). Table 1 summarizes the remarkable improvements from the point source scanning (scDCT) to line scanning approach (PS-DSCI) over the same ROI. Using the same camera with a frame rate of 24 fps (C11440-42U40, Hamamatsu) for both setups, the sampling rate increases from 0.0096 Hz (total 50 × 50 scanning points in scDCT) to 0.24 Hz (total 50 + 50 scanning lines in PS-DSCI), representing a ~25-fold improvement achieved by the PS-DSCI with total 100 scanning lines, used in the present study. Line scanning significantly reduces the number of raw intensity images acquired for blood flow reconstruction, leading to a 23-fold reduction in computation time and a 25-fold reduction in data storage. The larger the number of scanning points/lines, the higher the working efficiency.

**Table 1** Comparisons between the point scanning and line scanning

| Methods | Number of Images | Sampling Rate (Hz) | Computation Time (s) | Storage (Megabytes) |
|---|---|---|---|---|
| Point scanning | 2500 (50 × 50 points) | 0.0096 | ~ 410 | 20000 |
| Line scanning | 100 (50 + 50 lines) | 0.24 | ~ 17 | 800 |

*2.2 Data Analysis*

*2.2.1 Automatic Extraction of Detector Area/Belt Around Line-shape Source*

Our innovative approach for depth-sensitive blood flow reconstruction involves defining detector areas located at certain distances from the source center, with the goal of selectively capturing diffused photons originating from certain depths. The synchronization between the DMD and sCMOS camera ensures the capture of one intensity image at each source position. However, the line shapes of scanning sources typically transform into oval shapes due to the Gaussian distribution of light source and the curvature of the target tissue surface (Fig. 3a-3c and Fig. 3e-3g). We innovatively developed a new algorithm in MATLAB to extract the source properties at each scanning position and then used these properties to generate a detector area/belt with a predefined S-D separation (Fig. 3d and Fig. 3h).

Specifically, the captured intensity images were binarized and labeled, and morphological analyses were conducted using the "regionprops" function in MATLAB to identify the properties of oval-shaped sources. The modifiable elliptical-shape detector belts matching oval-shape sources were then defined by substituting the extracted properties into the equation of an ellipse (Eq. 1).

$$\frac{(x-h)^2}{a^2} + \frac{(y-k)^2}{b^2} = 1 \tag{1}$$

where $x$ and $y$ are rows and columns of the intensity image, respectively, $h$ and $k$ are the center of the oval-shape source, and a and b are major axis length or minor axis length, depending on the orientation of the source. These processes adaptively configured the detector belt, centered around the oval-shape source with the same S-D separation (Fig. 3d and Fig. 3h). The minor and major axis lengths (a and b) can be defined to adjust the S-D separation and detector-belt thickness.

*2.2.2 2D Mapping of Blood Flow Index (BFI)*

The diffuse laser speckle contrast ($K_s$) is calculated within the defined detector belt using Eq. 2 [47, 48]:

$$K_s = \frac{\sigma_s}{\langle I \rangle} = \frac{\sqrt{\langle I \rangle^2 - \langle I^2 \rangle}}{\langle I \rangle} \tag{2}$$

where $K_s$ is the ratio of standard deviation ($\sigma_s$) over mean intensity ($\langle I \rangle$) in an N × N (e.g., 3 × 3, 5 × 5, or 7 × 7) pixel window. A 2D matrix of $K_s$ values are then generated by aggregating $K_s$ values of all line scanning images while normalizing the overlapped detector areas. The BFI is approximated as BFI = $\frac{1}{K_s^2}$ [49, 50]. Following our previous method, convolution functions and kernel matrixes in MATLAB were used to improve computation efficiency [42]. Presently, the computation time for processing 100 scanning lines (Table 1) is ~17 seconds, which can be further reduced by GPU computing and leveraging MATLAB's Parallel Computation Toolbox [51].

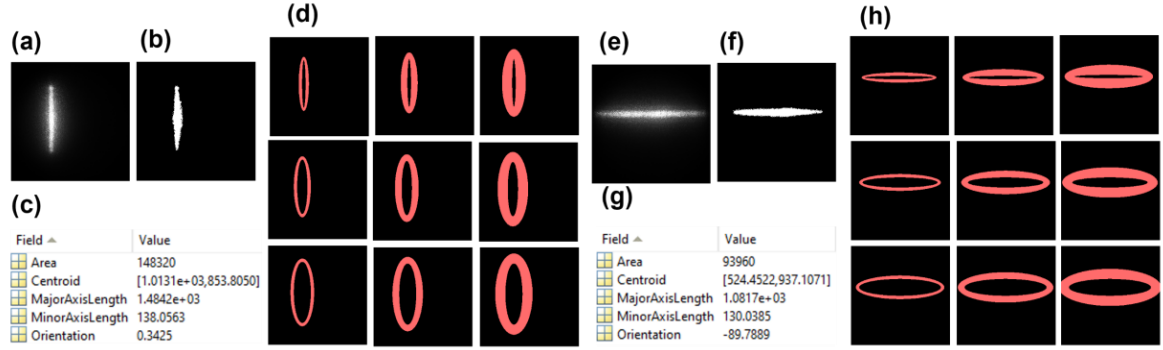

**Fig. 3** Extraction of detector area/belt around line-shape source. (a) and (e) Raw intensity images of vertical and horizontal oval-shape sources, respectively. (b) and (f) Binary masks of vertical and horizontal oval-shape sources, respectively. (c) and (g) An example of vertical and horizontal oval-shaped source properties extracted using the `regionprops` function in MATLAB's Image Processing Toolbox. (d) and (h) The elliptical-shape detector belts with varied S-D separations of 1 to 3 mm and varied detector-belt thicknesses of 1 to 3 mm.

*2.3 Head-simulating Phantom Experiments*

*2.3.1 Fabrication of Head-simulating Phantoms*

Standardized tissue-simulating phantoms with known optical properties and geometries are widely accepted for evaluating optical imaging technologies. Two solid phantoms with infinity-shape channels were printed (3D printer SL1, Prusa) with the top layer thicknesses of 1 mm and 2 mm to mimic the mouse skull with varied thicknesses [25]. Figure 4a-4c show the fabricated solid phantom, incorporating an infinity-shape channel with a diameter of 3.5 mm for the liquid phantom. The fabricated solid phantoms were filled with the liquid phantom solution and sealed by a thin layer of plastic and hot glue. The solid phantoms made of titanium dioxide ($TiO_2$), India ink (Black India, Massachusetts), and clear resin (eSUN Hard Tough) were designed to accommodate the liquid phantom solution composed of Intralipid solution (Fresenius Kabi, Sweden), India ink, and water. The concentration of India ink controlled the tissue absorption coefficient ($\mu_a$), while $TiO_2$ and Intralipid concentrations regulated the reduced scattering coefficient ($\mu_s'$). The optical properties of both the solid and liquid phantoms were set at $\mu_a =$ 0.03

cm$^{-1}$ and $\mu_s'$ = 9 cm$^{-1}$ [52]. The Brownian motion of Intralipid particles within the infinity-shape channel simulated particle flow to mimic the motion of red blood cells in vessels (i.e., blood flow).

*2.3.2 Experimental Setup and Procedures*

Each phantom was scanned with a total of 100 lines: 50 vertical and 50 horizontal lines. The camera collected 100 images on the ROI of 4 × 4 cm$^2$ at the working distance of 15 cm with the exposure time of 10 ms for each image and frame rate of 24 fps (equivalent sampling rate of 0.24 Hz). With an $f/\#$ of 8 and a magnification (M) of 0.32, the speckle size ($\rho_{speckle}$) is 20.2 µm. This speckle size is sufficiently larger than the camera pixel size (6.5 µm) to satisfy the Nyquist condition at 785 nm based on Eq. 3 [16]. Here, $M$ represents the magnification, $f/\#$ denotes the f-number of the zoom lens, and $\lambda$ is the laser wavelength.

$$\rho_{speckle} = 2.44(M + 1) * \lambda * f/\# \qquad (3)$$

*2.3.3 Signal-To-Noise Ratio (SNR) Analysis*

We experimentally evaluated the reconstructed flow imaging quality of head-simulating phantoms by analyzing the raw images with different pixel windows including 3 × 3, 5 × 5, 7 × 7, and 9 × 9. A window size of 5 × 5 was finally selected for calculating $K_s$ to balance between the spatial resolution and SNR. The detector-belt thickness and S-D separation varied from 1 to 6 mm respectively to investigate their impacts on SNRs of flow images. To quantify the SNR, a binary mask was created, representing the known infinity-shape flow channel. The SNR was calculated by taking the ratio of averaged flow values over the regions with Intralipid particle flow of the liquid phantom (valid flow signal, mask = 1) and without particle flow of the solid phantom outside the infinity-shape channel (noise, mask = 0).

## 2.4 In Vivo Experiments in Adult Mice

### 2.4.1 Experimental Setup and Procedures

All experimental procedures involving animals were approved by the University of Kentucky Institutional Animal Care and Use Committee (IACUC). Nine adult male mice aged from 9 to 18 weeks with different experimental procedures were imaged by the PS-DSCI (Table 2). The mouse was subjected to 1-2% Isoflurane anesthesia, and its hairs on the head and at cervical surgical site were removed by hair removal cream. The surgical skin was disinfected with Betadine followed by 70% Ethanol. The scalps of 8 mice (mouse #1 to mouse #8) were surgically removed to reduce partial volume effects of scalps on CBF maps. The head of one mouse (mouse #9) was kept intact (without scalp retraction) to test the capability of PS-DSCI penetrating through the intact head. The experimental setup was the same as that used for phantom experiments, except the ROI size of 3 × 3 cm². This ROI adjustment resulted in the magnification of 0.43 and speckle size of 21.9 µm, which still satisfied the Nyquist condition. 2D maps of BFI were reconstructed at different depths by calculating $K_s$ on a window size of 5 × 5. The relative time-course changes in CBF (rCBF) were calculated by normalizing BFI data to their baseline values before pathophysiological manipulations.

**Table 2** Subject information and experimental protocols

| Subjects | Age (Weeks) | Experimental Protocols | Scalp Retraction |
|---|---|---|---|
| Mouse #1 | 18 | 8% $CO_2$ inhalation, carotid artery ligation | Yes |
| Mouse #2 | 18 | 8% $CO_2$ inhalation, carotid artery ligation | Yes |
| Mouse #3 | 18 | 8% $CO_2$ inhalation, carotid artery ligation | Yes |
| Mouse #4 | 18 | 8% $CO_2$ inhalation, carotid artery ligation | Yes |
| Mouse #5 | 18 | 8% $CO_2$ inhalation, died during carotid artery ligation | Yes |
| Mouse #6 | 9 | 8% $CO_2$ inhalation, carotid artery ligation | Yes |
| Mouse #7 | 9 | 8% $CO_2$ inhalation, carotid artery ligation | Yes |
| Mouse #8 | 9 | 8% $CO_2$ inhalation, carotid artery ligation, 100% $CO_2$ inhalation | Yes |
| Mouse #9 | 9 | 8% $CO_2$ inhalation, carotid artery ligation | No |

*2.4.2 Continuous Imaging of Global rCBF in Mice During 8% $CO_2$ Inhalations*

$CO_2$ is a vasodilator, leading to a global increase in rCBF [53]. The mouse on a heating blanket was anesthetized (1-2% isoflurane) with its head secured on a stereotaxic frame. The PS-DSCI imaging was performed for 9 mice (8 mice with intact skull but retracted scalp, and mouse #9 with intact scalp and skull) before, during, and after a 5-minute exposure to a gas mixture consisting of 8% $CO_2$ and 92% $O_2$.

*2.4.3 Continuous Imaging of Regional rCBF in Mice During Sequential Transient Carotid Artery Ligations*

After recovering from 8% $CO_2$ inhalation, the mouse underwent sequential transient carotid arterial ligation surgeries to create sequential decreases in rCBF in the left hemisphere (LH) and right hemisphere (RH). Transient carotid arterial ligation surgery involves creating a midline incision to access both carotid arteries. A 6-0 braided nylon loose knot suture was placed around each carotid artery, allowing regional transient ligation by applying a mild pull on the corresponding knot [42, 54]. The PS-DSCI imaging was performed for 9 mice (8 mice with intact skull but retracted scalp, and mouse #9 with intact scalp and skull) during 5 minutes of baseline, 3 minutes of right carotid artery ligation, 1 minute of bilateral ligation, 3 minutes of left carotid artery release, and 5 minutes of right carotid artery release. At the end of the study, one mouse (mouse #8) was subjected to 100% $CO_2$ inhalation to induce a sharp reduction in rCBF.

*2.4.4 Statistical Analysis*

Statistical analysis was performed using a two-sided paired t-test in IBM SPSS Statistics software with the confidence interval of 95% to evaluate rCBF variations at different phases of 8% $CO_2$ inhalation and transient carotid artery ligations. The mean values of rCBF at different phases were calculated over all subjects and compared to their baseline values. A p-value of <0.05 was considered statistically significant for all statistical analysis.

## 3 Results

*3.1 PS-DSCI Enabled Depth-sensitive 2D Mapping of Particle Flow in Head Simulating Phantoms*

Figure 4 shows the fabricated head-simulating phantoms (Fig. 4a-4c), SNR analysis results (Fig. 4d and Fig. 4e), and reconstructed 2D flow maps (Fig. 4f and Fig. 4g). Figure 4d and Fig. 4e show SNR distributions with varied S-D separations and detector-belt thicknesses for the two phantoms with top layer thicknesses of 1 mm and 2 mm, respectively. For the phantom with a top layer thickness of 1 mm, higher SNR values appeared with the S-D separations of 3 mm to 5 mm and detector-belt thicknesses of 1 mm to 3 mm (Fig. 4d). For the phantom with a top layer thickness of 2 mm, higher SNR values shifted to larger S-D separations of 4 mm to 6 mm and thicker detector-belt thicknesses of 2 mm to 5 mm (Fig. 4e). Figure 4f shows the resulting 2D flow map of the phantom with a top layer thickness of 1 mm, reconstructed at the S-D separation of 4 mm and the detector-belt thickness of 2 mm. Similarly, Fig. 4g shows the resulting 2D flow map of the phantom with a top layer thickness of 2 mm, reconstructed at the S-D separation of 5 mm and the detector-belt thickness of 4 mm.

These results demonstrated the depth sensitivity of PS-DSCI measurements because larger S-D separations of 4 mm to 6 mm correspond to deeper imaging depths of 2 mm to 3 mm as compared to the smaller S-D separations of 3 mm to 5 mm at a shallower imaging depth of 1.5 mm to 2.5 mm. Deeper penetration and thicker top layer resulted in fewer diffused photons being detected, thus leading to lower SNRs. To conclude, our PS-DSCI enabled resolving the infinity-shape channel of Intralipid particle flow contrasts at the depths up to ~3 mm in the two phantoms with top layer thicknesses of 1 mm and 2 mm. These phantom experiments provide the guidance for optimizing S-D separations and detector-belt thicknesses to obtain the best image quality based on target head geometries.

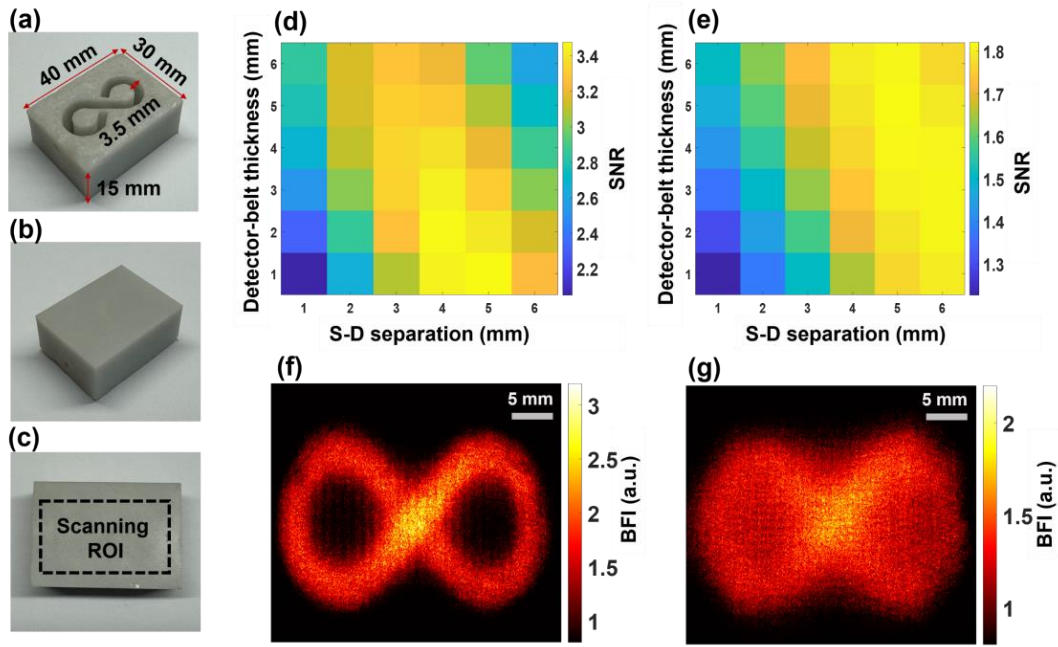

**Fig. 4** Depth-sensitive 2D mapping of Intralipid particle flow in head-simulating phantoms. (a) Top view of fabricated solid phantom with the infinity-shape channel, designed to be filled with the liquid phantom. (b) Bottom view of the fabricated phantom. (c) The selected ROI on the bottom of the phantom. (d) The SNR distribution with varied S-D separations (1 to 6 mm) and detector-belt thicknesses (1 to 6 mm) for the phantom with top layer thickness of 1 mm. (e) The SNR distribution with varied S-D separations (1 to 6 mm) and detector-belt thicknesses (1 to 6 mm) for the phantom with top layer thickness of 2 mm. (f) The 2D flow map of head-simulating phantom with the top layer of 1 mm thickness. The S-D separation and detector-belt thickness for flow map reconstruction were 4 mm and 2 mm, respectively. (g) The 2D flow map of head-simulating phantom with the top layer of 2 mm thickness. The S-D separation and detector-belt thickness for flow map reconstruction were 5 mm and 4 mm, respectively.

## 3.2 In Vivo Test Results

### 3.2.1 PS-DSCI Enabled 2D Mapping of BFI at Different Depths in Mice with Intact Skull

Figure 5 demonstrates the capability of PS-DSCI to generate 2D maps of BFI at different depths in one illustrative mouse (mouse #7). Because the average thickness of adult mouse skulls is about 0.5 mm to 1 mm [55] and based on the phantom test results shown in Fig. 4, 2D maps of BFI with varied S-D separations of 1 to 3 mm and detector-belt thicknesses of 1 to 3 mm are reconstructed and shown in Fig 5d. The results highlighted that when using smaller S-D separations and thinner detector belts, PS-DSCI generated BFI maps from superficial brain tissues with higher spatial resolutions and greater SNRs. Accordingly, the S-D separation and detector-belt thickness were set to 1 mm and 2 mm respectively for analyzing time-course rCBF changes in Fig. 6 and Fig. 7.

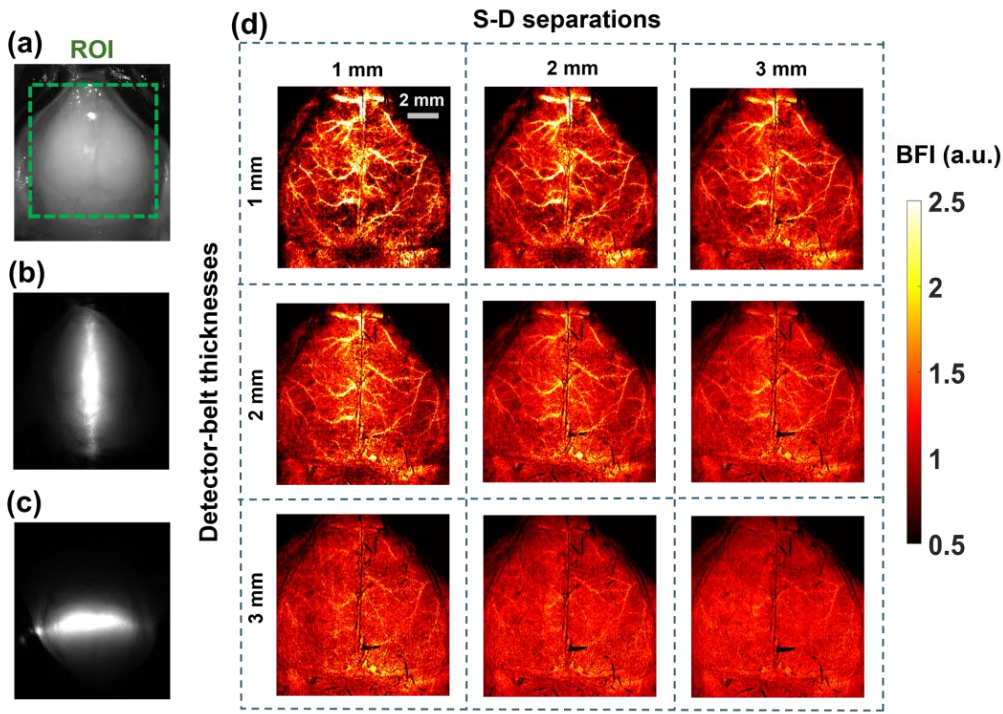

**Fig. 5** 2D maps of BFI at different depths in a representative mouse (#7). (a) The selected ROI for BFI mapping on the exposed skull with its scalp retracted. (b) The speckle image with a vertical scanning line on the ROI. (c) The speckle image with a horizontal scanning line on the ROI. (d) 2D maps of BFI with varied S-D separations of 1 to 3 mm and detector-belt thicknesses of 1 to 3 mm.

*3.2.2 PS-DSCI Enabled Continuous Mapping of Global rCBF Increases in Mice with Intact Skulls During 8% $CO_2$ Inhalations*

Figure 6a and Fig. 6b show time-course changes of rCBF before, during, and after 8% $CO_2$ inhalation in one illustrative mouse (mouse #8) and group mice (n = 8, mouse #1 to mouse #8,) with intact skulls, respectively. Table 3 summarizes group time-course changes in rCBF and corresponding p-values for comparing those changes at different phases of $CO_2$ inhalation relative to the baseline (assigning 100%). The average rCBF values at the baseline and recovery phases were quantified using the mean values of data during the entire baseline and recovery phases. The average rCBF values at the 8% $CO_2$ inhalation phase were quantified over a 2-minute period of the $CO_2$ inhalation phase, when peak rCBF increases were observed. The results show that the inhalation of 8% $CO_2$ resulted in a significant increase in rCBF (109.98% ± 2.26%, p = 0.003) from the baseline (100%), which lasted to the recovery phase (107.07% ± 2.36%, p = 0.02).

**Table 3** Group average rCBF changes (mean ± standard error) from their baselines (100%) during 8% $CO_2$ inhalations in 8 mice (mouse #1 to mouse #8)

| Brain region | 8% $CO_2$ inhalation | Recovery |
|---|---|---|
| Global rCBF | 109.98% ± 2.26%<br>p = 0.003 | 107.07% ± 2.36%<br>p = 0.02 |

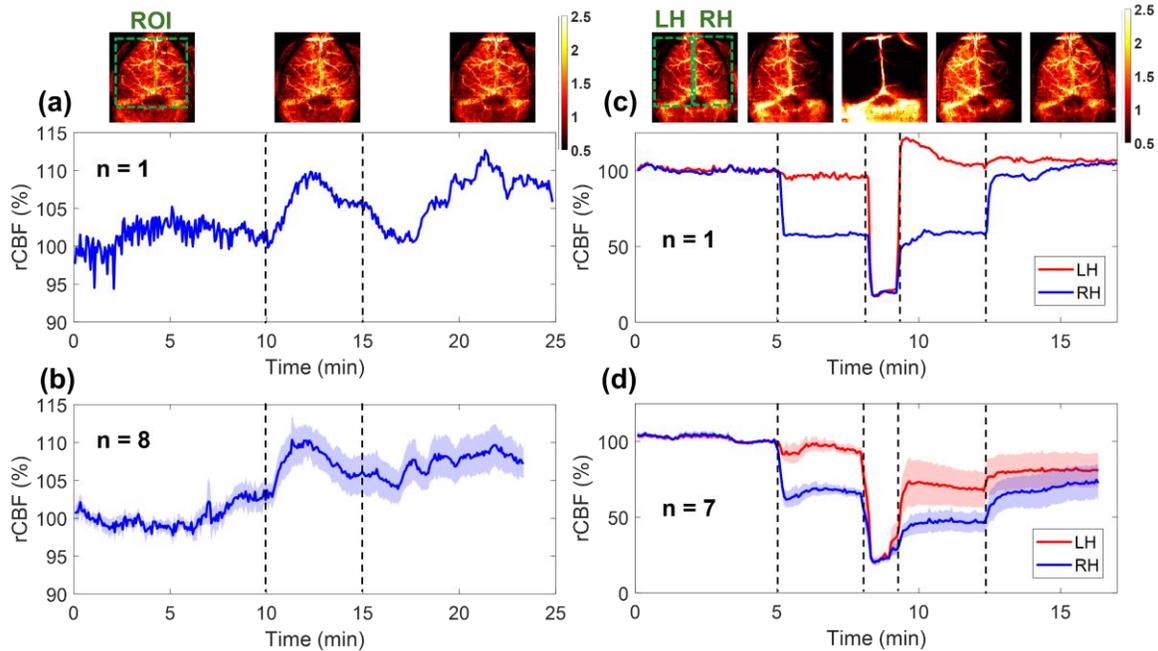

**Fig. 6** Continuous mapping of rCBF variations during 8% $CO_2$ inhalations and transient carotid artery ligations in mice with intact skulls. (a) BFI maps and time-course rCBF changes before, during, and after 8% $CO_2$ inhalation in an illustrative mouse (mouse #8). The dash lines separate 10 mins of baseline, 5 mins of $CO_2$ inhalation, and 10 mins of recovery, respectively. The ROI for data analysis of rCBF time-course changes is shown in the first BFI map. (b) Group average time-course rCBF changes during 8% $CO_2$ inhalations in 8 mice (mouse #1 to mouse #8). The error bars represent standard errors. (c) BFI maps and time-course rCBF changes during transient carotid artery ligations in an illustrative mouse (mouse #8). The dash lines separate the 5 mins of baseline, 3 mins of right carotid artery ligation, 1 min of bilateral ligation, 3 mins of left carotid artery release, and 5 mins of right carotid artery release. The ROI of LH and RH for data analyses of rCBF time-course changes are shown in the first BFI map. (d) Group average time-course rCBF changes during transient carotid artery ligations in 7 mice (mouse #1 to mouse #4 and mouse #6 to mouse #8). The error bars represent standard errors.

*3.2.3 PS-DSCI Enabled Continuous Mapping of Regional rCBF Reductions in Mice with Intact Skulls During Transient Carotid Artery Ligations*

After recovering from 8% $CO_2$ inhalation, 8 mice underwent sequential transient carotid arterial ligation surgeries. One mouse (mouse #5) was excluded from the group average results due to its unfortunate death after bilateral ligations. Figure 6c and Fig. 6d show time-course changes of rCBF at the LH and RH before, during, and after sequential transient carotid artery ligations in one illustrative mouse (mouse #8) and group mice (n = 7, excluding mouse #5), respectively.

Table 4 summarizes average changes in rCBF, along with the corresponding p-values for comparing these changes during different phases of transient carotid artery ligations to their respective baseline values. The mean rCBF values reported in Table 4 were quantified using the mean values of rCBF data during the entire periods of different phases. As expected, sequential ligations/releases of the right and left carotid arteries led to significant rCBF changes in relevant brain hemispheres, compared to their baselines.

Table 4 Group average rCBF changes (mean ± standard error) from their baselines (100%) during sequential transient carotid arterial ligations in 7 mice (excluding mouse #5)

| Brain region | Right ligation | Bilateral ligation | Release left ligation | Release right ligation |
| --- | --- | --- | --- | --- |
| RH rCBF | 66.58% ± 3.05% $p < 0.001$ | 23.62% ± 1.73% $p < 0.001$ | 44.01% ± 5.98% $p < 0.001$ | 64.70% ± 10.13% $p = 0.013$ |
| LH rCBF | 93.25% ± 2.89% $p = 0.058$ | 26.07% ± 1.49% $p < 0.001$ | 67.15% ± 1.25% $p = 0.039$ | 77.52% ± 11.16% $p = 0.091$ |

*3.2.4 PS-DSCI Enabled Continuous Mapping of Sharp rCBF Decrease in a Mouse with an Intact Skull During 100% $CO_2$ Inhalation*

Following the artery ligations/releases, one mouse (mouse #8) was subjected to 100% $CO_2$ inhalation to test the capability of PS-DSCI for capturing rapid rCBF decreases until its death. Figure 7 show time-course changes of rCBF before (assigned 100%) and during 100% $CO_2$ inhalation over ~5.5 minutes in mouse #8. A significant reduction in rCBF was observed immediately after inhaling 100% $CO_2$, followed by a transient increase at 3-4 minutes when the animal had long, deep, gasping breaths. Subsequently, rCBF steadily declined to approximately 5% of its baseline (100%) by the time when the animal was declared dead.

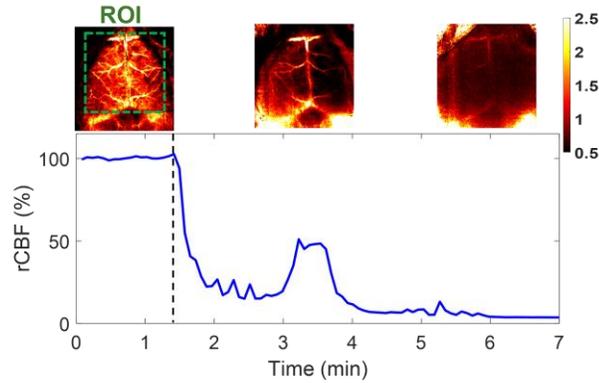

**Fig. 7** Continuous mapping of rCBF variations during 100% $CO_2$ inhalations in mouse #8 with an intact skull. BFI maps and time-course rCBF changes in mouse #8 before and during 100% $CO_2$ inhalation until its death. More than 95% decrease in rCBF was observed at the end of the experiment when mouse #8 died. The dash line separates the baseline and 100% $CO_2$ inhalation phases. The ROI for data analysis of rCBF time-course changes is shown in the first BFI map.

*3.2.5 PS-DSCI Enabled Noninvasive and Continuous Monitoring of rCBF Variations in a Mouse with Intact Head*

One mouse (mouse #9) with intact head (without scalp retraction) underwent 8% $CO_2$ inhalation and transient carotid artery ligations to evaluate the capability of PS-DSCI for noninvasive and continuous monitoring of rCBF variations. Given the increased total top layer thicknesses of scalp and skull, the S-D separation of 2 mm and detector-belt thickness of 2 mm were used for the reconstruction of BFI maps. Figure 8a shows the results from the continuous mapping of rCBF variations during 8% $CO_2$ inhalation in mouse #9. rCBF increased to 132.0% ± 2.7% and 135.2% ± 6.4% (mean ± standard deviation) at the maximum increase of 8% $CO_2$ inhalation phase and during the recovery phase from its baseline (100%), respectively. These rCBF changes during 8% $CO_2$ inhalation (Fig. 8a) were larger than those from the mice without scalp (Fig. 6a and Fig. 6b), likely due to the overlayed blood flow elevations from both the scalp and cortex in response to $CO_2$ in mouse #9.

The transient carotid artery ligations induced similar rCBF reductions in mouse #9 (Fig. 8b), compared to those from the mice without scalp (Fig 6c and Fig. 6d). We noted that the carotid artery ligations were not released completely in mouse #9, leading to incomplete rCBF recoveries in both hemispheres. Nevertheless, rCBF variations during artery ligations are clearly discernible. In addition, the partial volume effect of scalp rendered cerebral vasculatures in the reconstructed BFI maps.

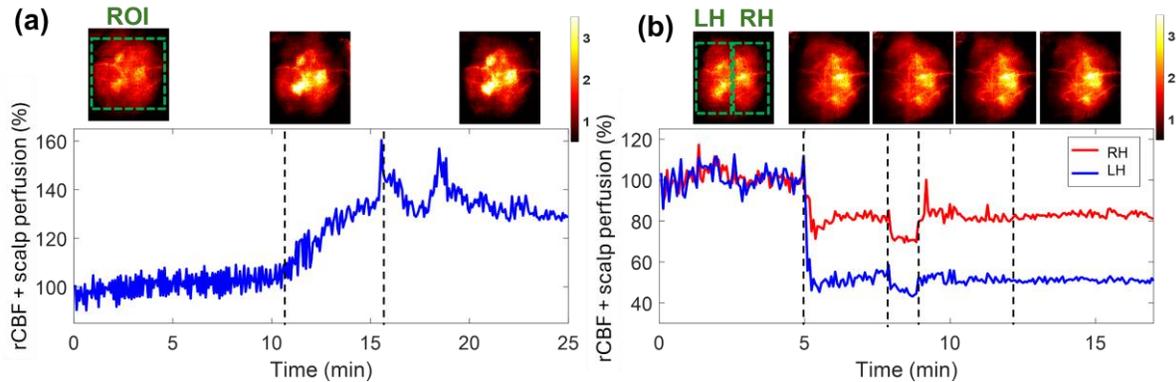

**Fig. 8** Continuous mapping of rCBF variations during 8% $CO_2$ inhalations and transient carotid artery ligations in mouse #9 with an intact head. (a) BFI maps and time-course rCBF changes before, during, and after 8% $CO_2$ inhalation in mouse with intact head. The dash lines separate 10 mins of baseline, 5 mins of $CO_2$ inhalation, and 10 mins of recovery, respectively. The ROI for data analysis of rCBF time-course changes is shown in the first BFI map. (b) BFI maps and time-course rCBF changes during transient carotid artery ligations in a mouse with intact head. The dash lines separate the 5 mins of baseline, 3 mins of left carotid artery ligation, 1 min of bilateral ligation, 3 mins of right carotid artery release, and 5 mins of left carotid artery release. The ROI of LH and RH for data analysis of rCBF time-course changes are shown in the first BFI map.

## 4 Discussion and Conclusions

The vitality and functionality of brain tissue are highly dependent on proper CBF circulation. Continuous imaging of CBF holds promise for the diagnosis and therapeutic management of many cerebral and neurovascular diseases. To meet such a critical need, we developed a DMD-based, line scanning, depth-sensitive PS-DSCI technique for noncontact and continuous imaging of CBF with remarkably improved spatiotemporal resolution (Fig. 1-3). While the line scanning can be achieved through various methods including a line laser [56] and a combination of a cylindrical lens and a galvo mirror [57], the programmable DMD maximizes the flexibility of operation. The DMD enables the creation of diverse and fast scanning patterns including the line shape (used in the present study) and point scanning to achieve an optimal spatiotemporal resolution. Compared to the point scanning in scDCT, the line scanning in PS-DSCI increases significantly the spatiotemporal resolution of CBF imaging and reduces the computation time and storage for image reconstructions (Table 1). The high sampling rate of PS-DSCI is crucial for capturing rapid CBF changes while high spatial resolution is important for visualizing brain vasculature. The combination of high spatiotemporal resolution holds the potential for depth-sensitivity mapping of brain FC, which is the subject of our future study.

In designing the PS-DSCI prototype (Fig. 1), the DMD was illuminated with the semi-collimated light to provide a scanning pattern with a large depth of focus, thus allowing for a flexible working distance [58, 59]. Although the use of collimated light limits the scanning ROI (e.g., $3 \times 1.5$ cm² in the existing design), PS-DSCI prototype enables CBF imaging in head-simulating phantoms and small rodents (Fig. 4-8). A plano-concave lens may be added in the projection path to enlarge the scanning ROI for future applications in larger subjects.

In analyzing collected PS-DSCI data, deformations of line shape sources were observed, resulting from the Gaussian distribution of light and curvature of target tissue surface. We innovatively developed new algorithms to extract the source properties at each scanning position. Based on the extracted source properties, elliptical shape detector areas/belts with predefined S-D separations were used to address deformations of line-shape scanning, thus minimizing BFI reconstruction artifacts (Fig. 3). The S-D separations and detector-belt thicknesses can be modified for reconstructing BFI images at different depths and with different SNRs.

To evaluate the depth sensitivity of PS-DSCI in mapping flow distributions at different depths, head-simulating phantoms with known optical and geometrical properties were fabricated and scanned (Fig. 4). Results demonstrate that with the optimized S-D separations and detector-belt thicknesses, our PS-DSCI enables resolving the infinity-shape channel of Intralipid particle flow contrasts at the depths up to ~3 mm in the two head-simulating phantoms with top layer thicknesses of 1 mm and 2 mm. These phantom experiments provide guidance to optimize the PS-DSCI for obtaining the best image quality in rodents with known head geometries (Fig. 5-8).

The capabilities of the PS-DSCI for 2D mapping of BFI at different depths and continuous monitoring of global and regional rCBF alterations were examined in 9 adult mice with intact skull (mouse #1 to mouse #8) or intact head (mouse #9) during a variety of pathophysiological manipulations (Table 2). Based on the variation in top layer thicknesses of the scalp/skull, the S-D separations and detector-belt thicknesses were optimized for the reconstruction of BFI maps at different depths (Fig. 5-8). The results in mice with intact skull but retracted scalp (mouse #1 to mouse #8) show that the inhalation of 8% $CO_2$ resulted in a significant global rCBF increases (109.98% ± 2.26%, Fig. 6b and Table 3) and sequential ligations/releases of the right and left carotid arteries led to significant regional rCBF variations in relevant brain hemispheres (Fig. 6d and Table 4). These results meet pathophysiological expectations and are generally consistent with previous studies in adult rodents using scDCT, LSCI, MRI, and PET and similar experimental protocols [42, 54, 60-64]. At the end of the experiment, mouse #8 was subjected to 100% $CO_2$ inhalation, when a remarkable reduction in rCBF was observed immediately (Fig. 7). Upon the animal's death, rCBF dropped to 5% of its baseline level (100%), demonstrating the sensitivity of PS-DSCI in detecting a minimal CBF level.

One mouse (mouse #9) with an intact head underwent 8% $CO_2$ inhalation and transient carotid artery ligations to evaluate the capability of PS-DSCI for noninvasive monitoring of rCBF variations. Results show larger rCBF changes during 8% $CO_2$ inhalation (Fig. 8a), compared to those from the mice without scalp (Fig. 6b), which is likely due to the overlayed blood flow elevations from both the scalp and cortex in response to $CO_2$ in mouse #9 with an intact head. The transient carotid artery ligations induced similar regional rCBF reductions in mouse #9 (Fig. 8b), compared to those from the mice without scalp (Fig. 6c and Fig. 6d). Even though the partial volume effect of scalp rendered cerebral vasculatures, rCBF variations during carotid artery ligations are clearly discernible.

In summary, to meet the need of fast, high-density, and continuous imaging of CBF at different depths, we have developed a low-cost, DMD-based, line scanning, depth-sensitive PS-DSCI technique. Compared to the point scanning in conventional scDCT, the line scanning in new PS-DSCI significantly increase spatiotemporal resolution of CBF imaging and reduces the computation time and storage for image reconstructions. New algorithms have been developed to address deformations of line-shape scanning, thus minimizing BFI reconstruction artifacts. The capabilities of the PS-DSCI for 2D mapping of BFI at different depths and continuous monitoring of global and regional rCBF alterations were examined in head-simulating phantoms and adult mice with known optical and geometrical properties of heads. The results from PS-DSCI are consistent with previous studies in adult rodents using other technologies and similar experimental protocols. Importantly, the high spatiotemporal resolution of PS-DSCI holds the promise for depth-sensitivity mapping of brain functional activity. Future studies will develop and optimize a low-cost, fast, mobile, and user-friendly PS-DSCI system with a greater tissue penetration depth, larger ROI, higher sampling rate, and lower computation time for cerebral functional imaging of larger subjects.

*Disclosures*

None.

*Codes and Data Availability*

All codes and data supporting the findings of this study are available from the corresponding author upon reasonable request.

*Acknowledgments*

We acknowledge the financial support from the National Institutes of Health (NIH) #R01 EB028792, #R01 HD101508, #R21 HD091118, #R21 NS114771, #R41 NS122722, #R42 MH135825, #R56 NS117587 (G. Y.); and Neuroscience Research Priority Area (NRPA) Pilot Grant from the University of Kentucky (L. C.). The content is solely the responsibility of the authors and does not necessarily represent the official views of NIH, NRPA, or University of Kentucky.

*References*

1. Zhu, M., J.J. Ackerman, and D.A. Yablonskiy, *Body and brain temperature coupling: the critical role of cerebral blood flow.* Journal of Comparative Physiology B, 2009. **179**: p. 701-710.

2. Willie, C.K., et al., *Integrative regulation of human brain blood flow.* The Journal of physiology, 2014. **592**(5): p. 841-859.

3. Claassen, J.A., et al., *Regulation of cerebral blood flow in humans: physiology and clinical implications of autoregulation.* Physiological reviews, 2021. **101**(4): p. 1487-1559.

4. Bandera, E., et al., *Cerebral blood flow threshold of ischemic penumbra and infarct core in acute ischemic stroke: a systematic review.* Stroke, 2006. **37**(5): p. 1334-1339.

5. Yonas, H., et al., *Increased stroke risk predicted by compromised cerebral blood flow reactivity.* Journal of neurosurgery, 1993. **79**(4): p. 483-489.

6. Campbell, B.C., et al., *Failure of collateral blood flow is associated with infarct growth in ischemic stroke.* Journal of Cerebral Blood Flow & Metabolism, 2013. **33**(8): p. 1168-1172.

7. Lin, P.-Y., et al., *Reduced cerebral blood flow and oxygen metabolism in extremely preterm neonates with low-grade germinal matrix-intraventricular hemorrhage.* Scientific reports, 2016. **6**(1): p. 25903.

8. Sunwoo, J., et al., *Diffuse correlation spectroscopy blood flow monitoring for intraventricular hemorrhage vulnerability in extremely low gestational age newborns.* Scientific reports, 2022. **12**(1): p. 12798.

9. Kim, J., et al., *Resting cerebral blood flow alterations in chronic traumatic brain injury: an arterial spin labeling perfusion FMRI study.* Journal of neurotrauma, 2010. **27**(8): p. 1399-1411.

10. Kenney, K., et al., *Cerebral vascular injury in traumatic brain injury.* Experimental neurology, 2016. **275**: p. 353-366.

11. Vartiainen, J., et al., *Functional magnetic resonance imaging blood oxygenation level-dependent signal and magnetoencephalography evoked responses yield different neural functionality in reading.* Journal of Neuroscience, 2011. **31**(3): p. 1048-1058.

12. Toronov, V., et al., *The roles of changes in deoxyhemoglobin concentration and regional cerebral blood volume in the fMRI BOLD signal.* Neuroimage, 2003. **19**(4): p. 1521-1531.

13. Van Zijl, P.C., et al., *Quantitative assessment of blood flow, blood volume and blood oxygenation effects in functional magnetic resonance imaging.* Nature medicine, 1998. **4**(2): p. 159-167.

14. Muzik, O., et al., *15O PET measurement of blood flow and oxygen consumption in cold-activated human brown fat.* Journal of Nuclear Medicine, 2013. **54**(4): p. 523-531.

15. Hiura, M., et al., *Changes in cerebral blood flow during steady-state cycling exercise: a study using oxygen-15-labeled water with PET.* Journal of Cerebral Blood Flow & Metabolism, 2014. **34**(3): p. 389-396.

16. Boas, D.A. and A.K. Dunn, *Laser speckle contrast imaging in biomedical optics.* Journal of biomedical optics, 2010. **15**(1): p. 011109-011109-12.

17. Senarathna, J., et al., *Laser speckle contrast imaging: theory, instrumentation and applications.* IEEE reviews in biomedical engineering, 2013. **6**: p. 99-110.

18. Shepherd, A.P. and P.Å. Öberg, *Laser-Doppler blood flowmetry*. Vol. 107. 2013: Springer Science & Business Media.

19. Fredriksson, I., M. Larsson, and T. Strömberg, *Measurement depth and volume in laser Doppler flowmetry.* Microvascular research, 2009. **78**(1): p. 4-13.

20. Boas, D.A., et al., *Establishing the diffuse correlation spectroscopy signal relationship with blood flow.* Neurophotonics, 2016. **3**(3): p. 031412-031412.

21. Yu, G., *Near-infrared diffuse correlation spectroscopy in cancer diagnosis and therapy monitoring.* Journal of biomedical optics, 2012. **17**(1): p. 010901-010901.

22. Durduran, T. and A.G. Yodh, *Diffuse correlation spectroscopy for non-invasive, micro-vascular cerebral blood flow measurement.* Neuroimage, 2014. **85**: p. 51-63.
25

23. Bi, R., J. Dong, and K. Lee, *Deep tissue flowmetry based on diffuse speckle contrast analysis.* Optics letters, 2013. **38**(9): p. 1401-1403.

24. Huang, C., et al., *Low-cost compact diffuse speckle contrast flowmeter using small laser diode and bare charge-coupled-device.* Journal of biomedical optics, 2016. **21**(8): p. 080501-080501.

25. Liu, X., et al., *Simultaneous measurements of tissue blood flow and oxygenation using a wearable fiber-free optical sensor.* Journal of Biomedical Optics, 2021. **26**(1): p. 012705-012705.

26. Lin, Y., et al., *Three-dimensional flow contrast imaging of deep tissue using noncontact diffuse correlation tomography.* Applied Physics Letters, 2014. **104**(12).

27. Allen, J. and K. Howell, *Microvascular imaging: techniques and opportunities for clinical physiological measurements.* Physiological measurement, 2014. **35**(7): p. R91.

28. Dragojević, T., et al., *High-density speckle contrast optical tomography of cerebral blood flow response to functional stimuli in the rodent brain.* Neurophotonics, 2019. **6**(4): p. 045001-045001.

29. Yu, G., Y. Lin, and C. Huang, *Noncontact three-dimensional diffuse optical imaging of deep tissue blood flow distribution.* US Patent #9/861,319, 2018.

30. Huang, C., et al., *Noncontact 3-D Speckle Contrast Diffuse Correlation Tomography of Tissue Blood Flow Distribution.* IEEE Trans Med Imaging, 2017. **36**(10): p. 2068-2076.

31. Mazdeyasna, S., et al., *Noncontact speckle contrast diffuse correlation tomography of blood flow distributions in tissues with arbitrary geometries.* J Biomed Opt, 2018. **23**(9): p. 1-9.

32. Huang, C., et al., *Noncontact diffuse optical assessment of blood flow changes in head and neck free tissue transfer flaps.* J Biomed Opt, 2015. **20**(7): p. 75008.

33. Huang, C., et al., *Noninvasive noncontact speckle contrast diffuse correlation tomography of cerebral blood flow in rats.* Neuroimage, 2019. **198**: p. 160-169.

34. Bonaroti, A., et al., *The Role of Intraoperative Laser Speckle Imaging in Reducing Postoperative Complications in Breast Reconstruction.* Plast Reconstr Surg, 2019. **144**(5): p. 933e-934e.

35. Huang, C., et al., *Speckle contrast diffuse correlation tomography of cerebral blood flow in perinatal disease model of neonatal piglets.* J Biophotonics, 2021. **14**(4): p. e202000366.

36. Irwin, D., et al., *Near-infrared Speckle Contrast Diffuse Correlation Tomography (scDCT) for Noncontact Imaging of Tissue Blood Flow Distribution*. 2022: CRC Press.

37. Hamedi, F., et al. *An affordable miniaturized speckle contrast diffuse correlation tomography (scDCT) device for 2D mapping of cerebral blood flow*. in *Multiscale Imaging and Spectroscopy V*. 2024. SPIE.

38. Akbari, F., et al. *Laser coherence length requirement for speckle contrast diffuse correlation tomography (scDCT)*. in *Multiscale Imaging and Spectroscopy V*. 2024. SPIE.

39. Saikia, M.J. and R. Kanhirodan, *Region-of-interest diffuse optical tomography system.* Review of Scientific Instruments, 2016. **87**(1).

40. Han, G., et al., *Optimization of source-detector separation for non-invasive regional cerebral blood flow sensing.* Infrared Physics & Technology, 2021. **117**: p. 103843.

41. Strangman, G.E., Z. Li, and Q. Zhang, *Depth sensitivity and source-detector separations for near infrared spectroscopy based on the Colin27 brain template.* PloS one, 2013. **8**(8): p. e66319.

42. Mohtasebi, M., et al., *Depth-sensitive diffuse speckle contrast topography for high-density mapping of cerebral blood flow in rodents.* Neurophotonics, 2023. **10**(4): p. 045007-045007.

43. Bauer, A.Q., et al., *Optical imaging of disrupted functional connectivity following ischemic stroke in mice.* Neuroimage, 2014. **99**: p. 388-401.

44. White, B.R., et al., *Bedside optical imaging of occipital resting-state functional connectivity in neonates.* Neuroimage, 2012. **59**(3): p. 2529-2538.

45. Yoshida, Y., M. Nakao, and N. Katayama, *Resting-state functional connectivity analysis of the mouse brain using intrinsic optical signal imaging of cerebral blood volume dynamics.* Physiological measurement, 2018. **39**(5): p. 054003.

46. Mohtasebi, M., et al., *Detection of low-frequency oscillations in neonatal piglets with speckle contrast diffuse correlation tomography.* Journal of Biomedical Optics, 2023. **28**(12): p. 121204-121204.

47. Fercher, A.F. and J.D. Briers, *Flow Visualization by Means of Single-Exposure Speckle Photography.* Optics Communications, 1981. **37**(5): p. 326-330.

48. Dunn, A. and W.J. Tom, *Methods of producing laser speckle contrast images*. 2014, Google Patents.

49. Sunil, S., et al., *Guidelines for obtaining an absolute blood flow index with laser speckle contrast imaging.* bioRxiv, 2021: p. 2021.04. 02.438198.

50. Postnikov, E.B., M.O. Tsoy, and D.E. Postnov. *Matlab for laser speckle contrast analysis (lasca): a practice-based approach*. in *Saratov Fall Meeting 2017: Laser Physics and Photonics XVIII; and Computational Biophysics and Analysis of Biomedical Data IV*. 2018. SPIE.

51. Singh, D., et al., *A fast algorithm towards real-time laser speckle contrast imaging.* J. Biomed. Opt, 2022. **15**: p. 011109.

52. Barnett, A.H., et al., *Robust inference of baseline optical properties of the human head with three-dimensional segmentation from magnetic resonance imaging.* Applied optics, 2003. **42**(16): p. 3095-3108.

53. Tancredi, F.B. and R.D. Hoge, *Comparison of cerebral vascular reactivity measures obtained using breath-holding and $CO_2$ inhalation.* Journal of Cerebral Blood Flow & Metabolism, 2013. **33**(7): p. 1066-1074.

54. Fathi, F., et al., *Time-resolved laser speckle contrast imaging (TR-LSCI) of cerebral blood flow.* arXiv preprint arXiv:2309.13527, 2023.

55. Nguyen, T., et al., *Repeated closed-head mild traumatic brain injury-induced inflammation is associated with nociceptive sensitization.* Journal of Neuroinflammation, 2023. **20**(1): p. 196.

56. Maity, A.K., et al., *SpeckleCam: high-resolution computational speckle contrast tomography for deep blood flow imaging.* Biomedical Optics Express, 2023. **14**(10): p. 5316-5337.

57. Du, E., et al., *Line scan spatial speckle contrast imaging and its application in blood flow imaging.* Applied Sciences, 2021. **11**(22): p. 10969.

58. Xiong, Z., et al., *Illumination uniformity improvement in digital micromirror device based scanning photolithography system.* Optics Express, 2018. **26**(14): p. 18597-18607.

59. Hellman, B. and Y. Takashima, *Angular and spatial light modulation by single digital micromirror device for multi-image output and nearly-doubled étendue.* Optics Express, 2019. **27**(15): p. 21477-21496.

60. Struys, T., et al., *In vivo evidence for long-term vascular remodeling resulting from chronic cerebral hypoperfusion in mice.* Journal of Cerebral Blood Flow & Metabolism, 2017. **37**(2): p. 726-739.

61. Choki, J., et al., *Effect of carotid artery ligation on regional cerebral blood flow in normotensive and spontaneously hypertensive rats.* Stroke, 1977. **8**(3): p. 374-379.

62. Ewing, J.R., et al., *Direct comparison of local cerebral blood flow rates measured by MRI arterial spin-tagging and quantitative autoradiography in a rat model of experimental cerebral ischemia.* Journal of Cerebral Blood Flow & Metabolism, 2003. **23**(2): p. 198-209.

63. Ohshima, M., et al., *Cerebral blood flow during reperfusion predicts later brain damage in a mouse and a rat model of neonatal hypoxic–ischemic encephalopathy.* Experimental neurology, 2012. **233**(1): p. 481-489.

64. Seker, F.B., et al., *Neurovascular reactivity in the aging mouse brain assessed by laser speckle contrast imaging and 2-photon microscopy: quantification by an investigator-independent analysis tool.* Frontiers in Neurology, 2021. **12**: p. 745770.
29

**Caption List**

**Fig. 1** PS-DSCI system. (a) The distribution of diffraction orders and energy envelope reflected by the DMD, which are dependent on the angle of incidence ($\theta i$) and angle of reflection ($\theta r$). $\theta r$ represents the center of energy envelope distribution. (b) Schematic of the PS-DSCI prototype. (c) A photo of PS-DSCI prototype.

**Fig. 2** Point scanning (scDCT) versus line scanning (PS-DSCI). In contrast to 2500 scanning points (S1 to S2500) by scDCT, PS-DSCI scans the same ROI with 100 scanning lines (L1 to L50, along vertical and horizontal directions, respectively).

**Fig. 3** Extraction of detector area/belt around line-shape source. (a) and (e) Raw intensity images of vertical and horizontal oval-shape sources, respectively. (b) and (f) Binary masks of vertical and horizontal oval-shape sources, respectively. (c) and (g) An example of vertical and horizontal oval-shaped source properties extracted using the `regionprops` function in MATLAB's Image Processing Toolbox. (d) and (h) The elliptical-shape detector belts with varied S-D separations of 1 to 3 mm and varied detector-belt thicknesses of 1 to 3 mm.

**Fig. 4** Depth-sensitive 2D mapping of Intralipid particle flow in head-simulating phantoms. (a) Top view of fabricated solid phantom with the infinity-shape channel, designed to be filled with the liquid phantom. (b) Bottom view of the fabricated phantom. (c) The selected ROI on the bottom of the phantom. (d) The SNR distribution with varied S-D separations (1 to 6 mm) and detector-belt thicknesses (1 to 6 mm) for the phantom with top layer thickness of 1 mm. (e) The SNR distribution with varied S-D separations (1 to 6 mm) and detector-belt thicknesses (1 to 6 mm) for the phantom with top layer thickness of 2 mm. (f) The 2D flow map of head-simulating phantom with the top layer of 1 mm thickness. The S-D separation and detector-belt thickness for flow map reconstruction were 4 mm and 2 mm, respectively. (g) The 2D flow map of head-simulating

phantom with the top layer of 2 mm thickness. The S-D separation and detector-belt thickness for flow map reconstruction were 5 mm and 4 mm, respectively.

**Fig. 5** 2D maps of BFI at different depths in a representative mouse (#7). (a) The selected ROI for BFI mapping on the exposed skull with its scalp retracted. (b) The speckle image with a vertical scanning line on the ROI. (c) The speckle image with a horizontal scanning line on the ROI. (d) 2D maps of BFI with varied S-D separations of 1 to 3 mm and detector-belt thicknesses of 1 to 3 mm.

**Fig. 6** Continuous mapping of rCBF variations during 8% $CO_2$ inhalations and transient carotid artery ligations in mice with intact skulls. (a) BFI maps and time-course rCBF changes before, during, and after 8% $CO_2$ inhalation in an illustrative mouse (mouse #8). The dash lines separate 10 mins of baseline, 5 mins of $CO_2$ inhalation, and 10 mins of recovery, respectively. The ROI for data analysis of rCBF time-course changes is shown in the first BFI map. (b) Group average time-course rCBF changes during 8% $CO_2$ inhalations in 8 mice (mouse #1 to mouse #8). The error bars represent standard errors. (c) BFI maps and time-course rCBF changes during transient carotid artery ligations in an illustrative mouse (mouse #8). The dash lines separate the 5 mins of baseline, 3 mins of right carotid artery ligation, 1 min of bilateral ligation, 3 mins of left carotid artery release, and 5 mins of right carotid artery release. The ROI of LH and RH for data analyses of rCBF time-course changes are shown in the first BFI map. (d) Group average time-course rCBF changes during transient carotid artery ligations in 7 mice (mouse #1 to mouse #4 and mouse #6 to mouse #8). The error bars represent standard errors.

**Fig. 7** Continuous mapping of rCBF variations during 100% $CO_2$ inhalations in mouse #8 with an intact skull. BFI maps and time-course rCBF changes in mouse #8 before and during 100% $CO_2$ inhalation until its death. More than 95% decrease in rCBF was observed at the end of the experiment when mouse #8 died. The dash line separates the baseline and 100% $CO_2$ inhalation phases. The ROI for data analysis of rCBF time-course changes is shown in the first BFI map.

**Fig. 8** Continuous mapping of rCBF variations during 8% $CO_2$ inhalations and transient carotid artery ligations in mouse #9 with an intact head. (a) BFI maps and time-course rCBF changes before, during, and after 8% $CO_2$ inhalation in mouse with intact head. The dash lines separate 10 mins of baseline, 5 mins of $CO_2$ inhalation, and 10 mins of recovery, respectively. The ROI for data analysis of rCBF time-course changes is shown in the first BFI map. (b) BFI maps and time-course rCBF changes during transient carotid artery ligations in a mouse with intact head. The dash lines separate the 5 mins of baseline, 3 mins of left carotid artery ligation, 1 min of bilateral ligation, 3 mins of right carotid artery release, and 5 mins of left carotid artery release. The ROI of LH and RH for data analysis of rCBF time-course changes are shown in the first BFI map.

**Table 2** Comparisons between the point scanning and line scanning

**Table 2** Subject information and experimental protocols

**Table 3** Group average rCBF changes (mean ± standard error) from their baselines (100%) during 8% $CO_2$ inhalations in 8 mice (mouse #1 to mouse #8)

**Table 4** Group average rCBF changes (mean ± standard error) from their baselines (100%) during sequential transient carotid arterial ligations in 7 mice (excluding mouse #5)